\documentclass{emulateapj}

\usepackage{graphics,epsfig}

\shorttitle{Support of neutrals against gravity in prominences}
\shortauthors{Terradas et al.}

\begin{document}

\title{On the support of neutrals against gravity in solar prominences}

\author{J. Terradas, R. Soler, R. Oliver, \& J. L., Ballester} 
\affil{Departament de F\'\i sica, Universitat de les Illes Balears, E-07122
Palma de Mallorca, Spain} \email{jaume.terradas@uib.es}

\begin{abstract} Cool and dense prominences found in the solar atmosphere are
known to be partially ionized because of their relative low temperature. In this
Letter, we address the long-standing problem of how the neutral component of the
plasma in prominences is supported against gravity. Using the multiple fluid
approach we solve the time-dependent equations in two dimensions considering the
frictional coupling between the neutral and ionized components of the magnetized
plasma representative of a solar prominence embedded in a hot coronal
environment. We demonstrate that given an initial density enhancement in the two
fluids, representing the body of the prominence, the system is able to relax in
the vicinity of magnetic dips to a stationary state in which both neutrals and
ionized species are dynamically suspended above the photosphere. Two different
coupling processes are considered in this study, collisions between ions and
neutrals and charge exchange interactions. We find that for realistic conditions
ions are essentially static while neutrals have a very small downflow velocity.
The coupling between ions and neutrals is so strong at the prominence body that
the behavior is similar to that of a single fluid with an effective density equal
to the sum of the ion and neutral species. We also find that the charge exchange
mechanism is about three times more efficient sustaining neutrals than elastic
scattering of ions with neutrals.\end{abstract}



\maketitle

\section{Introduction}

Traditionally the problem of solar prominence support is described in terms of
the magnetic force that balances the solar gravitational force. In the existing
models it is assumed that the prominence plasma is fully ionized. However,
observations of the prominence body in $\rm H\alpha$, which is an excitation
line, suggest that the observed plasma can not be fully ionized. The same happens
for prominence observations in the \ion{He}{1}  $10830\,\rm \AA$ line. At a
temperature of $20,000\,\rm K$, typical of the prominence corona transition region
(PCTR), we already have ionized $\rm H$, ionized $\rm He$ and also ionized $\rm
Ca$. Thus, the support of both the ionized and neutral components is required.
The main problem is that the neutral component is not directly affected by the
restoring magnetic forces that can balance gravity. 

In the past it has been proposed that the frictional coupling between neutrals
and ions through elastic collisions might play a role in the support of
prominences.  \citet{mercierheyvaerts1977} examined the relative downward
diffusion of neutral atoms due to gravity and taking into account the magnetic
field. These authors found that the resulting downward velocity is irrelevant to
explain the mass loss in prominences. \citet{bakharevaetal1992} did the first
attempt to construct a model based on the one-dimensional Kippenhahn-Schl\"uter
solutions including ion-neutral collisions and provided some details about the
dynamical regimes of prominence evolution. \citet{pecengvold2000} suggested that
wave damping caused predominantly by ion-neutral collisions in the prominence
core may balance the acceleration of gravity. \citet{gilbertetal2002} have shown
that the draining effect for a hydrogen-helium plasma is rather small, specially
for the hydrogen that moves down in the direction of the photosphere at a typical
velocity of only $3.7\, \rm m \,s^{-1}$. The results of these works indicate that
ion-neutral coupling may be quite relevant to sustain neutrals, nevertheless, a
consistent study taking into account frictional coupling, magnetic forces, and
gravity has not been developed so far.

Another mechanism that provides a frictional force is the resonant charge
exchange process. Contrary to the elastic scattering between ions and neutrals,
charge exchange collisions are not identity-preserving. For hydrogen an energetic
proton captures the electron from a lower-energy neutral producing an energetic
neutral which has essentially the same energy as the incident proton. The reader
is referred to \citet{goldston1995} for details about this process.
\citet{leakeetal2013} has shown that under chromospheric conditions this process
increases the collisional coupling between ions and neutrals. It is therefore
logical to think that this mechanism can be also relevant under prominence
conditions.

In the present Letter we follow an approach similar to the one described in
\citet{terradasetal2013} for a fully ionized plasma and is based on solving the
time-dependent problem in a two-dimensional geometry with a magnetic field that
incorporates dips. The main difference is in the injection of a plasma dominated
by neutrals at the core of the prominence. With the two-fluid approximation the
effects of ion-neutral collisions and charge exchange collisions are properly
incorporated to the model. These mechanisms provide a redistribution of momentum
of the species and change the force balance in the system. This is the crucial
step to accomplish a new stationary state that is in dynamical equilibrium. 


\section{Basic two-fluid equations} 

We use the most simplified version of the two-fluid equations including
ion-neutral collisions and charge exchange collisions. For the ion-electron fluid we have the equations of
continuity, momentum, pressure, and magnetic induction,
\begin{eqnarray}
\frac{\partial{\rho_{\rm i}}}{\partial t}&=&-\nabla \cdot \left({\rho_{\rm i} \bf
v_i}\right),\label{eq1} \\
\frac{\partial{\left(\rho_{\rm i} \bf  v_i\right)}}{\partial t}&=&-\nabla \cdot \left(\rho_{\rm i} {\bf v_i} {\bf v_i} +p_{\rm ie}{\bf I} -{\bf
B}{\bf B} +\frac{1}{2}{B}^2{\bf I}\right)\nonumber \\
 & &+\rho_{\rm i} {\bf g}-\alpha_{\rm in}\left({\bf v_i}-{\bf
 v_n}\right)-\beta^{\rm cx}_{\rm in}\left({\bf v_i}-{\bf v_n}\right),\label{eq2}\\
 \frac{\partial{p_{\rm ie}}}{\partial t}&=&-\left(\bf v_i \cdot \nabla\right) {p_{\rm ie}}
-\gamma p_{\rm ie} \nabla \cdot {\bf v_i}\nonumber \\
& &+\left(\gamma-1\right)\left(W_{\rm in}+W^{\rm cx}_{\rm in}\right),\label{eq3}
\\ \frac{\partial{\bf B}}{\partial t}&=&\nabla\times\left(\bf v_i \times \bf
B\right),\label{eq4}
\end{eqnarray}
\noindent We have used the simplest version of the Ohm's law in which Ohmic
resistivity, the Hall and battery terms are ignored since they are very small
under typical prominence conditions.

For the fluid composed of neutrals we have
\begin{eqnarray}
 \frac{\partial{\rho_{\rm n}}}{\partial t}&=&-\nabla \cdot \left({\rho_{\rm n} \bf
 v_n}\right), \label{eq5}\\
\frac{\partial{\left(\rho_{\rm n} \bf  v_n\right)}}{\partial t}&=&-\nabla \cdot \left(\rho_{\rm n} {\bf v_n} {\bf v_n} +p_{\rm n}{\bf I}
\right)\nonumber \\
& &+{\rho_{\rm n} \bf g}+\alpha_{\rm in}\left({\bf v_i}-{\bf
v_n}\right)+\beta^{\rm cx}_{\rm in}\left({\bf v_i}-{\bf v_n}\right),\label{eq6}\\
 \frac{\partial{p_{\rm n}}}{\partial t}&=&-\left(\bf v_n \cdot \nabla\right) {p_{\rm n}}
-\gamma p_{\rm n} \nabla \cdot {\bf v_n}\nonumber \\
& &+\left(\gamma-1\right)\left(W_{\rm ni}+W^{\rm cx}_{\rm ni}\right).\label{eq7}
\end{eqnarray}

\noindent In these equations all the variables have the usual meaning. We define
the drift velocity as ${\bf v_{\rm D}}\equiv {\bf v_i}-{\bf v_n}$. The coupling
between the two fluids is through the parameters $\alpha_{\rm in}$ and
$\beta^{\rm cx}_{\rm in}$. The first parameter represents the friction
coefficient due to collisions between ions and neutrals
\citep[e.g.,][]{braginskii1965,chapmancowling1970}.  For hydrogen and assuming
that ions and electrons have the same temperature the friction coefficient is
given by the following expression \begin{eqnarray}\label{alph} \alpha_{\rm
in}={\rho_{\rm i}} {\rho_{\rm n}}\frac{1}{2 m_{\rm p}}\sigma_{\rm
in}\sqrt{\frac{4}{\pi} \left(v_{\rm Tie}^2+v_{\rm Tn}^2\right)}, \end{eqnarray}
\noindent where $m_{\rm p}$ is the proton mass and $\sigma_{\rm in}$ is the
momentum transfer cross section, taken to be equal to $10^{-18} \rm m^{2}$ in
this work \citep[see][]{vranjeskrstic2013}. The thermal speed of the different
species is given by $v_{\rm T}=\sqrt{2 k_B T/m_{\rm p}}$ where $T$ is the
temperature. It is  useful to define the collision frequency between neutrals and
ions as $\nu_{\rm ni}=\alpha_{\rm in}/\rho_{\rm n}$. The maximum collision
frequency in our configuration is $\nu_{\rm ni}=118\,\rm Hz$.

The second parameter, $\beta^{\rm cx}_{\rm in}$, is responsible for the charge
exchange collisions and is given by the following expression
\citep[see][]{paulsetal1995,meier2011}
\begin{eqnarray}\label{beta}
\beta^{\rm cx}_{\rm in}={\rho_{\rm i}} {\rho_{\rm n}}\frac{1}{
m_{\rm p}}& &\sigma_{\rm cx}
\left(\sqrt{\frac{4}{\pi}
\left(v_{\rm Tie}^2+v_{\rm Tn}^2\right)+v_{\rm D}^2}\right. \nonumber\\  
& &\left.+\frac{v_{\rm Tn}^2}{\sqrt{4
\left(\frac{4}{\pi} v_{\rm Tie}^2+v_{\rm D}^2\right)+\frac{9\pi}{4}v_{\rm
Tn}^2}}\right. \nonumber \\ 
& & \left.+\frac{v_{\rm Tie}^2}{\sqrt{4
\left(\frac{4}{\pi} v_{\rm Tn}^2+v_{\rm D}^2\right)+\frac{9\pi}{4}v_{\rm Tie}^2}}
\right).
\end{eqnarray}
For interactions between protons and neutral hydrogen the cross
section, $\sigma_{\rm cx}$,  is around $10^{-18} \rm m^{2}$
\citep[see][]{meiershumlak2012}. Hence, the
cross sections associated to elastic collisions and charge exchange collisions
are approximately the same.

The terms $W_{\rm in}$, $W_{\rm ni}$, $W^{\rm cx}_{\rm in}$, and $W^{\rm cx}_{\rm
ni}$ in the pressure equations contain frictional heating and thermal transfer
between the two populations associated to the two frictional mechanisms. The
explicit form of  $W_{\rm in}=-W_{\rm ni}$ can be found in, for example,
\citet{draine1986} or \citet{leakeetal2014}. The expressions for $W^{\rm cx}_{\rm
in}=-W^{\rm cx}_{\rm ni}$,  which are complex because the charge exchange
physics is more complicated, can be found in the works of \citet{meier2011} and
\citet{leakeetal2012}. Because of the thermal transfer between the species
included in these terms the temperatures of ions and neutrals will tend to be
equal on short time scales.

Equations~(\ref{alph})-(\ref{beta}) indicate that the friction coefficients
depend essentially on the density and pressure of the individual species (i.e.,
the thermal speed). Since the coefficients depend on the local plasma parameters
they also vary in space and time during the evolution,  allowing us to model
self-consistently the coupling between ions and neutrals.

From Eqs.~(\ref{eq2}) and (\ref{eq6}) we have that the force between the fluids
is proportional to the difference in velocity between ions and neutrals, ${\bf
v_{\rm D}}$. Static stationary solutions are simply not possible.
If neutrals are sustained against gravity it means that there must be a flow
that provides the required frictional restoring force. The question is whether
the flow produces an important draining effect of neutrals from the prominence
body.

\section{Initial conditions and method}

In our model the density of ions and neutrals is constructed using a static
background plus a localized enhancement representing the prominence. For the
background we assume a stratified atmosphere for ions with a scale height
characteristic of $10^{6}\,\rm K$, with gravity ($g=274\, \rm m\, s^{-2}$) 
pointing in the negative $z-$direction.

The prominence is represented by a large density enhancement respect to the
background \citep[see][]{terradasetal2013}, imposing that its maximum is 100
times larger than the coronal density. This enhancement is assumed to be
composed by $75\%$ of neutrals and $25\%$ of ions. These percentages are
consistent with the ionization models of \citet{gouttelabrosse2009} (see middle
panel of their Figure~1). Regarding the spatial structure of the enhancements
for both ions and neutrals is given by a two-dimensional Gaussian with a
characteristic width $a$ and $b$ in the $x$ and $z-$directions, respectively. We
use that $a=0.3\,L_0$ and $b=0.6\,L_0$. 

For the magnetic configuration we use a force-free quadrupolar structure
described in \citet{terradasetal2013}. Magnetic dips are present in this
topology (see Figure~\ref{figure2}) and are important to achieve a sustained
prominence. A small shear component is included in the model to avoid the
plasma-$\beta$ to be infinite at the base of the corona. We use a magnetic field
of $50\,\rm G$ at this level which gives $10\,\rm G$ at the core of the
prominence (the plasma-$\beta$ is around 0.1) and the changes in the
magnetic field during the evolution are rather small.

The system of time-dependent nonlinear equations given by
Eqs.~(\ref{eq1})-(\ref{eq7}) together with the initial conditions are solved in
two-dimensions. We choose the following normalizing values for the length,
$L_0=10^{7}\, \rm m$, and density, $\rho_0=5.2\times 10^{-13}\, \rm kg\,
m^{-3}$. The normalizing value for velocity is $c_{s0}=1.66\times 10^{5}\rm m\,
s ^{-1}$. The simulation domain extends from $-6L_0$ to $6L_0$ in the
$x-$direction and from $0$ to $8L_0$ in the $z-$direction. We use $200\times200$
grid points. Line-tying conditions are imposed at $z=0$ representing the base of
the corona while extrapolated conditions are enforced at the lateral and upper
boundaries. The inclusion of the frictional terms imposes a strong limitation on
the time steps required to properly resolve the short diffusion time scales.

\section{Results}

A representative simulation with the parameters given in the previous section is
described here. For a better understanding of the results we first neglect charge
exchange collisions, which will be included later, and concentrate on elastic
scattering between ions and neutrals.

\subsection{Ion-neutral collisions}

\begin{figure}[!h] \center{\includegraphics[width=9cm]{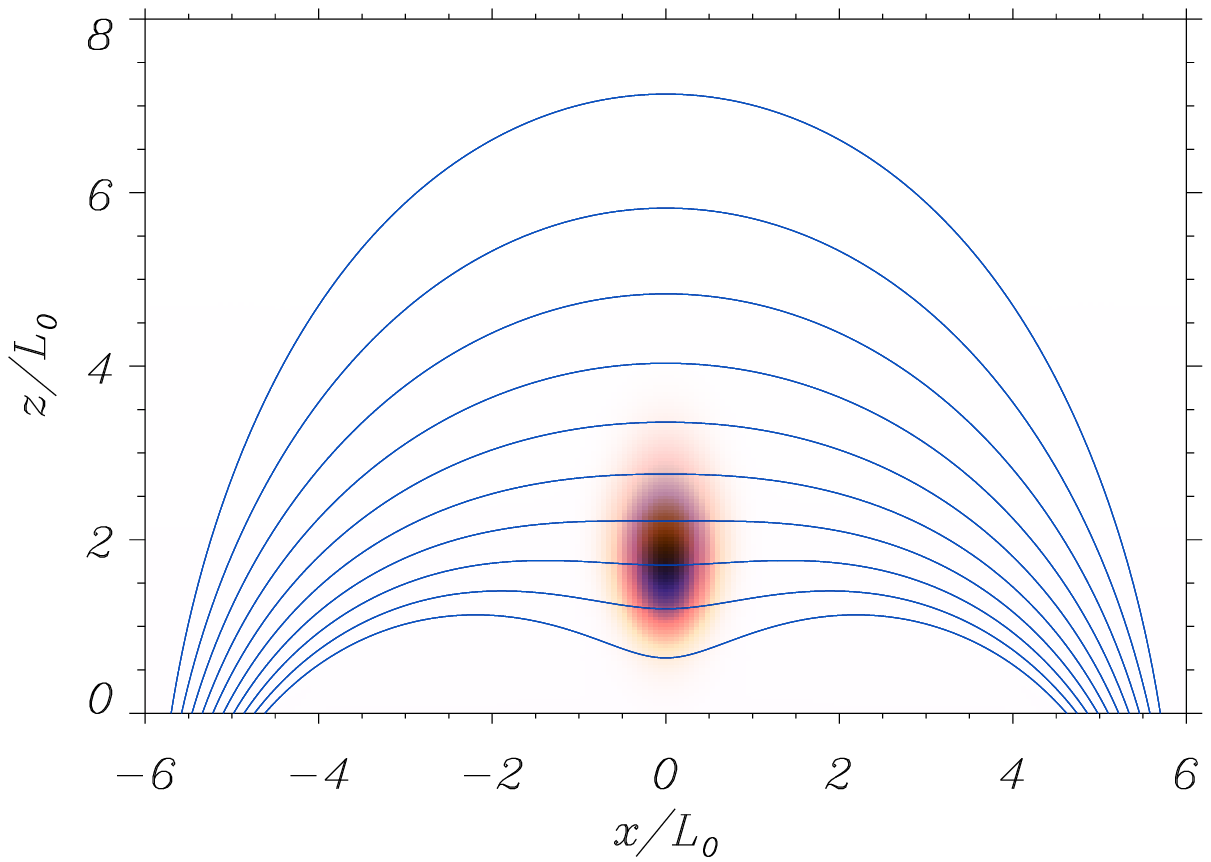}} 
\center{\includegraphics[width=9cm]{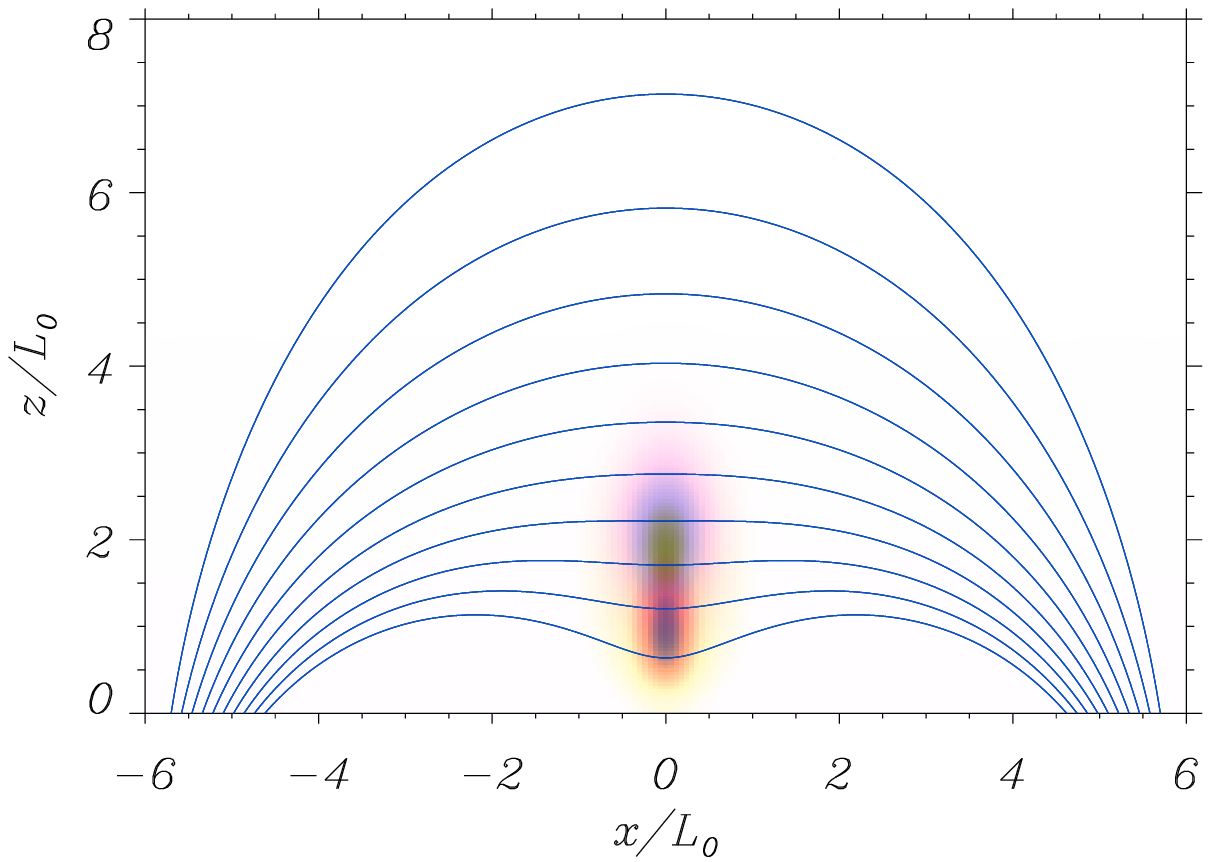}}  \caption{\small Magnetic field
(blue lines) and density of the ion-neutral fluid (blue-green colors) and
neutrals (yellow-pink colors) at a given time ($t=4.8\,\rm min$). In the top panel the simulation
corresponds to a maximum collision frequency of $\nu_{\rm ni}=8.5\,\rm Hz$,
approaching to the strong coupled case. For the bottom panel $\nu_{\rm
ni}=0.8\,\rm Hz$, the weak coupling between the ionized plasma and neutrals
produces the diffusion of neutrals across the magnetic field.}\label{figure2}
\end{figure}

From the analysis of the time-dependent simulations we find several evidences
about the support of neutrals located at the core of the prominence. As an
example, we have plotted in Figure~\ref{figure2} top panel, the two-dimensional
distribution of the two plasma densities after only $4.8\, \rm min$ of evolution.
The ion-neutral fluid and neutrals are essentially superimposed, and the
compact spatial shape of the densities found in this figure is very similar to
that at the beginning of the simulation meaning that the mass redistribution is
rather small in the configuration. On the contrary, in Figure~\ref{figure2}
bottom panel, the results of the same simulation are shown but with a reduced
maximum value for $\alpha_{\rm in}$ (ten times smaller than in top panel). The
behavior of the system is very different with respect to the previous case. Neutrals
are falling quite quickly and diffusing across the magnetic field. A sustained
situation for neutrals is not achieved in this case. 


We return to the case studied in Figure~\ref{figure2} top panel. We have allowed
the system to evolve for more than $10\,\rm hours$, and the geometrical changes
found at the prominence core are very small. In fact neutrals are moving
downwards but with such small velocity that makes, to all practical purposes,
difficult to distinguish from a static situation.  These results clearly
indicate that a sustained partially ionized plasma is a feasible configuration.
The crucial point is that the friction coefficient is very large at the core of
the prominence and the coupling between ions and neutrals is very strong.
Essentially the multicomponent plasma behaves as a single fluid at the
prominence body and this means that the  velocities for the two plasma
components are very similar. Thus, the drift velocity is rather small, but since
the frictional coefficient  is very large the frictional force, which is simply
the product $\alpha_{\rm in} {\bf v_{\rm D}}$, is sufficient to balance gravity.
This drift velocity, ${\bf v_{\rm D}}$, is always positive and therefore the 
frictional force counterbalances the gravity force pointing downwards and acting
on neutrals. The gas pressure at the center of the  prominence has a maximum,
and the pressure gradient is zero. The first term on the right hand side of
Eq.~(\ref{eq6}) is therefore negligible since $\rho_{\rm n} v_{{\rm n}z}^{2} $
is quite small (we concentrate only on the $z-$component). The remaining terms
provide the force balance in the $z-$direction, meaning that $v_{ {\rm
D}z}\approx g/ \nu_{\rm ni}$ in agreement with the results of
\citet{gilbertetal2002,gilbert2011} based on simple assumptions and the balance
in the momentum equation. This means that the drift velocity is inversely
proportional to the neutral-ion frequency. So if the system is in a stationary
state under balance of forces the stronger the coupling between the two-fluids
(higher neutral-ion frequency) the smaller the vertical component of the drift
velocity. In fact, the ionized plasma is in this regime essentially static ($v_{
{\rm i}z}\approx 0$), meaning that the downward velocity of neutrals is simply
$v_{ {\rm n}z}\approx -g/ \nu_{\rm ni}$.

Now we turn out attention to the ionized plasma and consider Eq.~(\ref{eq2}).
For the same reasons as before the gradient of the inertial and pressure terms
is negligible but the magnetic term is very relevant. It provides the balance
against the gravity term and the frictional force that now is pointing
downwards. To have force balance the magnetic force must be equal to  
$\rho_{\rm i} {g}+\alpha_{\rm in}\, v_{{\rm D}z}$, which using the previously
estimated value for $v_{{\rm D}z}$, it reduces to the simple and illustrative
expression $g\left(\rho_{\rm i}+\rho_{\rm n}\right)$. The physical
interpretation is clear. The ionized fluid is always supported by the magnetic
field but if a neutral component is included in the model the only way to have a
new balance of forces is to increase the restoring magnetic force in such a way
that the weight of the two plasma components is compensated by the deformation
of the magnetic field. This result is only valid when the two-fluids are
strongly coupled, and this is true at the core of the prominence body.

\begin{figure}[!h] \center{\includegraphics[width=9.5cm]{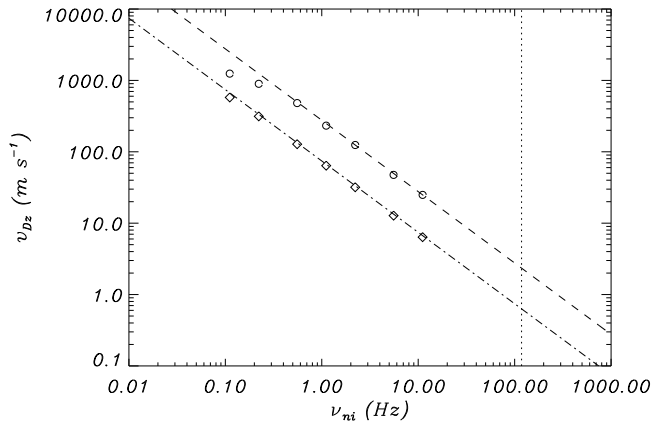}} 
\caption{\small Drift velocity for different values of $\alpha_{\rm in}$
($\nu_{\rm ni}$) inferred from the simulations at the point $x=0$, $z=2 L_0$.
Circles correspond to elastic ion-neutral collisions while diamonds to elastic
plus charge exchange collisions. The dashed line represent to the analytic
approximation $v_{ {\rm D}z}\approx g/\nu_{\rm ni}$, while the dot-dashed line
corresponds to the  approximation $v_{{\rm D}z}\approx g/(3.72\,\nu_{\rm ni})$. The
vertical dotted line is the maximum collision frequency. }\label{figureg}
\end{figure}

We have explored the dependence of the results on the crucial magnitude of our
model, namely  $\alpha_{\rm in}$ or equivalently $\nu_{\rm ni}$. We have
artificially imposed the maximum of this parameter and have performed different
numerical experiments. The results are shown in Figure~\ref{figureg}. When 
$\nu_{\rm ni}$ is around three orders of magnitude below the maximum value given
by Eq.~(\ref{alph}) ($\nu_{\rm ni}=118\,\rm Hz$), neutrals are weakly coupled
with the ion-electron fluid and they essentially fall down rapidly toward the
base of the corona (see Figure~\ref{figure2} bottom panel). However, when this parameter
is raised  a new dynamical  equilibrium is accomplished with a reduced drift
velocity. We have been able to check using the numerical results that the
expression $v_{ {\rm n}z}\approx -g/ \nu_{\rm ni}$ is satisfied (compare circles
with the dashed line), which is a further confirmation that the system is under a
dynamical balance. In fact, due to numerical issues related to the size of the
time-step in the simulations when the collision frequency is large, we have been
forced to fix a maximum collision frequency below the maximum value inferred from
Eq.~(\ref{alph}). Nevertheless from Figure~\ref{figureg} we see that we are
approaching to the analytic results. 

Using the maximum collision frequency ($\nu_{\rm ni}=118\,\rm Hz$) in our
configuration, and the expression for the velocity drift, we obtain that the
draining of neutrals takes place at a speed of only $v_{ {\rm n}z0}\approx -2.3\,
\rm m \,s^{-1}$ (see the dotted vertical line in Figure~\ref{figureg}). This is a
very low velocity and is in good agreement with the value inferred by
\citet{gilbertetal2002} ($v_{ {\rm n}z0}\approx -3.7\, \rm m \,s^{-1}$). Hence,
the draining of neutrals from the prominence core proceeds on very large
time-scales which are of the order of months using the simple time estimate
$2\,L_0/|v_{ {\rm n}z0}|$ (the time required for the neutrals located at a height
$2\,L_0$ to reach the base of the corona moving at a constant velocity $|v_{ {\rm
n}z0}|$).

In the two-fluid approach the heating rate due to the conversion of kinetic
energy is simply $\alpha_{\rm in} {\bf v_D}^{2}$. Using the expression for the
drift velocity we obtain $\rho_{\rm n} g^{2} /\nu_{\rm ni}$. This magnitude at
the core of the prominence (using the maximum frequency of $118\,\rm Hz$) gives a
heating rate which is typically of the order of $10^{-8}\,\rm J\, m^{-3}\,
s^{-1}$. Hence, heating by frictional collisions is quite irrelevant.

\subsection{Charge exchange collisions}

Now the terms associated to charge exchange interactions are activated in the
simulations. These terms increase the frictional coupling between the species,
and this can be easily inferred from the comparison of Eqs.~(\ref{alph}) and
(\ref{beta}). Neglecting the drift velocity in front of the thermal velocities,
and assuming that the thermal velocities of the ion-electron and neutral fluids
are the same, it is straight forward to find that $\beta^{\rm cx}_{\rm in}\approx
2.72\,\alpha_{\rm in}$ (since the two processes have the same cross section).
Thus, charge exchange collisions produce almost three times more friction than
that of ion-neutral collisions. This means that the downward velocity of neutrals
is further reduced by charge exchange interactions and the final expression is
$v_{ {\rm n}z}\approx -g \rho_{\rm n}/\left(\alpha_{\rm in}+ \beta^{\rm cx}_{\rm
in}\right)\approx -g/ \left(3.72\, \nu_{\rm ni}\right)$. To test the assumptions
made to derive this expression we have again changed the maximum collision
frequency and have calculated from the simulations the numerical value of $v_{
{\rm n}z}$ when both ion-neutral and charge exchange collisions are present. The
results are overplotted in Figure~\ref{figureg} with diamonds. Once more the
agreement between the simulations and the analytic approximation (plotted with a
dot-dashed line) is remarkable.

\section{Discussion}

In spite of the simplicity of our model it contains the very basic ingredients to
demonstrate that the support of neutrals in solar prominences through frictional
coupling is a viable mechanism. We have not tried to model the formation process
and have concentrated mainly on the issue of the support of neutrals. We have
found that as long as the friction coefficient is large, and the estimations of
this coefficient under prominence conditions point in this direction, it is
relatively easy to find a ``dynamical equilibrium". We have shown for the first
time by solving the time-dependent problem  that through the frictional coupling,
magnetic forces rearrange to balance the gravity force acting on the joint mass
of neutrals and ions.  Interestingly, for hydrogen the friction coefficient for
change exchange collisions is three times larger than the friction coefficient
for ion-neutral collisions. The nonstatic equilibrium is characterized by a
rather small downflow velocity  for neutrals, around $2.3\, \rm m \,s^{-1}$, for
ion-neutral collisions. But when charge exchange collisions are added to
ion-neutral collisions this downflow velocity is reduced up to a factor four,
since the net friction is increased. Therefore, the draining of neutrals from the
prominence is not an important issue when the frictional coupling between the
species is strong. To all practical purposes, it would be difficult to
distinguish a static situation from the steady-state situation obtained in this
work with such low values of the drift velocity.

Many effects have been ignored in the present work. In particular, the
ionization fraction does not change consistently because photoionization and
recombination have been neglected. It is relatively simple to include
recombination in the two-fluid equations but photoionization is much more
difficult. To properly address this problem the full NLTE (non-local
thermodynamic equilibrium) radiative transport equations should be coupled with
the two-fluid equations and this is out of the scope of this work. Viscosity and
other non-ideal effects have been neglected. It has also been assumed that
the prominence is an isolated system, more representative of quiescent
prominences, ignoring dynamical effects such as strong flows along the magnetic
field that are often observed in active region prominences which could modify
the draining rate of neutrals.

Finally, instead of 2D,  3D prominences models should be studied, since are able
to develop several kinds of instabilities such as the magnetic Rayleigh-Taylor
instability \citep[see][]{hillieretal11,terradasetal2015}. The effect of
neutrals on this instability needs to be investigated further since the
growth-rates can be significantly modified
\citep[see][]{diazetal2012,khomenkoetal2014}. 

\acknowledgements \small J.T. acknowledges support from the Spanish Ministerio
de Educaci\'on y Ciencia through a Ram\'on y Cajal grant. R.S. acknowledges
support from MINECO through a Juan de la Cierva grant (JCI-2012-13594), from
MECD through project CEF11-0012, and from the Vicerectorat d'Investigaci\'o i
Postgrau of the UIB. The authors acknowledge the funding provided under the
project AYA2011-22846 by the Spanish MICINN and FEDER Funds. The authors also
thank P. Heinzel and the anonymous referee for useful comments.


\begin{thebibliography}{22}
\expandafter\ifx\csname natexlab\endcsname\relax\def\natexlab#1{#1}\fi

\bibitem[{{Bakhareva} {et~al.}(1992){Bakhareva}, {Zaitsev}, \&
  {Khodachenko}}]{bakharevaetal1992}
{Bakhareva}, N.~M., {Zaitsev}, V.~V., \& {Khodachenko}, M.~L. 1992, \solphys,
  139, 299

\bibitem[{{Braginskii}(1965)}]{braginskii1965}
{Braginskii}, S.~I. 1965, Reviews of Plasma Physics, 1, 205

\bibitem[{{Chapman} \& {Cowling}(1970)}]{chapmancowling1970}
{Chapman}, S., \& {Cowling}, T.~G. 1970, {The mathematical theory of
  non-uniform gases. an account of the kinetic theory of viscosity, thermal
  conduction and diffusion in gases}

\bibitem[{{D{\'{\i}}az} {et~al.}(2012){D{\'{\i}}az}, {Soler}, \&
  {Ballester}}]{diazetal2012}
{D{\'{\i}}az}, A.~J., {Soler}, R., \& {Ballester}, J.~L. 2012, \apj, 754, 41

\bibitem[{{Draine}(1986)}]{draine1986}
{Draine}, B.~T. 1986, \mnras, 220, 133

\bibitem[{{Gilbert}(2011)}]{gilbert2011}
{Gilbert}, H. 2011, in American Institute of Physics Conference Series, ed.
  V.~{Florinski}, J.~{Heerikhuisen}, G.~P. {Zank}, \& D.~L. {Gallagher}, Vol.
  1366, 5--12

\bibitem[{{Gilbert} {et~al.}(2002){Gilbert}, {Hansteen}, \&
  {Holzer}}]{gilbertetal2002}
{Gilbert}, H.~R., {Hansteen}, V.~H., \& {Holzer}, T.~E. 2002, \apj, 577, 464

\bibitem[{Goldston \& Rutherford(1995)}]{goldston1995}
Goldston, R., \& Rutherford, P. 1995, Introduction to Plasma Physics (Institute
  of Physics Pub)

\bibitem[{{Gouttebroze} \& {Labrosse}(2009)}]{gouttelabrosse2009}
{Gouttebroze}, P., \& {Labrosse}, N. 2009, \aap, 503, 663

\bibitem[{{Hillier} {et~al.}(2011){Hillier}, {Isobe}, {Shibata}, \&
  {Berger}}]{hillieretal11}
{Hillier}, A., {Isobe}, H., {Shibata}, K., \& {Berger}, T. 2011, \apjl, 736, L1

\bibitem[{{Khomenko} {et~al.}(2014){Khomenko}, {D{\'{\i}}az}, {de Vicente},
  {Collados}, \& {Luna}}]{khomenkoetal2014}
{Khomenko}, E., {D{\'{\i}}az}, A., {de Vicente}, A., {Collados}, M., \& {Luna},
  M. 2014, \aap, 565, A45

\bibitem[{{Leake} {et~al.}(2013){Leake}, {Lukin}, \& {Linton}}]{leakeetal2013}
{Leake}, J.~E., {Lukin}, V.~S., \& {Linton}, M.~G. 2013, Physics of Plasmas,
  20, 061202

\bibitem[{{Leake} {et~al.}(2012){Leake}, {Lukin}, {Linton}, \&
  {Meier}}]{leakeetal2012}
{Leake}, J.~E., {Lukin}, V.~S., {Linton}, M.~G., \& {Meier}, E.~T. 2012, \apj,
  760, 109

\bibitem[{{Leake} {et~al.}(2014){Leake}, {DeVore}, {Thayer}, {Burns},
  {Crowley}, {Gilbert}, {Huba}, {Krall}, {Linton}, {Lukin}, \&
  {Wang}}]{leakeetal2014}
{Leake}, J.~E., {et~al.} 2014, \ssr, 184, 107

\bibitem[{{Meier}(2011)}]{meier2011}
{Meier}, E.~T. 2011, PhD thesis, University of Washington

\bibitem[{{Meier} \& {Shumlak}(2012)}]{meiershumlak2012}
{Meier}, E.~T., \& {Shumlak}, U. 2012, Physics of Plasmas, 19, 072508

\bibitem[{{Mercier} \& {Heyvaerts}(1977)}]{mercierheyvaerts1977}
{Mercier}, C., \& {Heyvaerts}, J. 1977, \aap, 61, 685

\bibitem[{{Pauls} {et~al.}(1995){Pauls}, {Zank}, \& {Williams}}]{paulsetal1995}
{Pauls}, H.~L., {Zank}, G.~P., \& {Williams}, L.~L. 1995, \jgr, 100, 21595

\bibitem[{{P{\'e}cseli} \& {Engvold}(2000)}]{pecengvold2000}
{P{\'e}cseli}, H., \& {Engvold}, O. 2000, \solphys, 194, 73

\bibitem[{{Terradas} {et~al.}(2013){Terradas}, {Soler}, {D{\'{\i}}az},
  {Oliver}, \& {Ballester}}]{terradasetal2013}
{Terradas}, J., {Soler}, R., {D{\'{\i}}az}, A.~J., {Oliver}, R., \&
  {Ballester}, J.~L. 2013, \apj, 778, 49

\bibitem[{{Terradas} {et~al.}(2015){Terradas}, {Soler}, {Luna}, {Oliver}, \&
  {Ballester}}]{terradasetal2015}
{Terradas}, J., {Soler}, R., {Luna}, M., {Oliver}, R., \& {Ballester}, J.~L.
  2015, \apj, 799, 94

\bibitem[{{Vranjes} \& {Krstic}(2013)}]{vranjeskrstic2013}
{Vranjes}, J., \& {Krstic}, P.~S. 2013, \aap, 554, A22

\end{thebibliography}
\end{document}